# AUTOMATIC ELECTRIC METER READING SYSTEM: A COST-FEASIBLE ALTERNATIVE APPROACH IN METER READING FOR BANGLADESH PERSPECTIVE USING LOW-COST DIGITAL WATTMETER AND WIMAX TECHNOLOGY

TANVIR AHMED[1], MD SUZAN MIAH[2], MD. MANIRUL ISLAM[3] and MD. RAKIB UDDIN[4]

## ABSTRACT

Energy meter reading is a monotonous and an expensive task. Now the meter reader people goes to each meter and take the meter reading manually to issue the bill which will later be entered in the billing software for billing and payment automation. If the manual meter reading and bill data entry process can be automated then it would reduced the laborious task and financial wastage. "Automatic Electric Meter Reading (AMR) System" is a metering system that is to be used for data collecting from the meter and processing the collected data for billing and other decision purposes. In this paper we have proposed an automatic meter reading system which is low cost, high performance, highest data rate, highest coverage area and most appropriate for Bangladesh perspective. In this AMR system there are four basic units. They are reading unit, communication unit, data receiving and processing unit and billing system. For reading unit we identified the disk rotation of the energy meter and stored the data in microcontroller. So it is not required to change the current analog energy meter. An external module will be added with the current energy meter. In the communication unit Wimax transceiver was used for wireless communication between meter end and the server end because of its wide coverage area. In the data receiving and processing unit meter reading will be collected from the transceiver which is controlled by another microcontroller. There will be a computer application that will take the data from the microcontroller. This will also help to avoid any tampering or break down of energy meter. There are various AMR system exists all over the world. Those systems were analyzed and we found they are not feasible for Bangladesh. Our proposed system is completely new and is appropriate for Bangladesh perspective. The study was conducted at the Electrical Circuit Laboratory, American International University, Dhaka, Bangladesh during October 2009 to November 2010.

**Keywords:** Digital wattmeter, AMR, WiMax and PLM.

## INTRODUCTRION

Automatic Meter Reading System (AMR) is the remote collection of consumption data from customers' utility like Electric meters using radio frequency, telephony, power-line or satellite communications technologies and process the data to generate the bill (Moghavvemi *et al.,* 2005). Now a day, AMR is heavily used in the abroad for collecting reading and billing purpose. In Bangladesh the electricity billing system is completely manual. The electric meters are situated in the houses, offices and factories etc .Meter readers go to the place which are generally situated inside the house and take the meter reading. Most of the time the owner gives some extra money to the meter reader person to have less meter reading. As a result corruptions occur and actual payment is not received by the service provider. So the provider faces a huge amount of loss in every year. At this present situation, it's very necessary to implement AMR in Bangladesh. Millions of Analog meters are already used in our houses, offices, industries. So we are not proposed any new meter. We want to develop a miniature module which will take reading from analog meter and then convert this data into digital data. And this module also responsible for transmission of data to the provider end. At the provider end there will be another module which is responsible for data receiving. And this module makes an interface with computer which is responsible for data processing. Automatic meter reading, or AMR, is the technology of automatically collecting data from energy meter or water metering devices (water, gas, and electric) and transferring that data to a central database for billing and/or analyzing. This means that billing can be based on actual consumption rather than on an estimate based on previous consumption, giving customers better control of their use of electric energy, gas usage, or water consumption (http://www.romdev-systems.com/index.php?page=ecm-definition). Automatic meter reading was first tested 30 years ago when trials were conducted by AT&T in cooperation with a group of utility companies and Westinghouse. After those successful experiments, AT&T offered to provide phone system-based Automatic Meter Reading services at $2 per meter the price was four times more than the

---
[1]Department of Computer Science, American International University-Bangladesh (AIUB) Dhaka, Email: tanvir@aiub.edu,
[2]Department of Electronics and Information System, Politecnico Di Torino Turin, Italy, Email : sznmiah@gmail.com,
[3]Department of Computer Science, American International University-Bangladesh (AIUB) Dhaka, Email: manirul@aiub.edu and
[4]Department of Pharmaceutical Chemistry, University of Dhaka, Dhaka, Bangladesh, Email : rakibuddin.md@gmail.com.

monthly cost of a person to read the meter-50 cents. Thus the program was considered economically unfeasible (http://www.usclcorp.com/news/Automatic_Power_Reading.pdf). The modern era of automatic meter reading began in 1985, when several major full-scale projects were implemented. Hackensack Water Co. and Equitable Gas Co. were the first to commit to full-scale implementation of automatic meter reading on water and gas meters, respectively. In 1986, Minnegasco initiated a 450,000- point radio-based automatic meter reading system. In 1987, Philadelphia Electric Co., faced with a large number of inaccessible meters, installed thousands of distribution line carrier automatic meter reading units to solve this problem. Thus, automatic meter reading is becoming more viable each day. Advances in solid-state electronics, microprocessor components and low-cost surface-mount technology assembly techniques have been the catalyst to produce reliable cost-effective products capable of providing the economics and human benefits that justify automatic meter reading systems on a large, if not full-scale basis [10]. AMR technologies include handheld, mobile and network technologies based on telephony platforms (wired and wireless), radio frequency (RF), or power line transmission (Agustín Zaballos *et al.,* 2009; John Newbury and William Miller, 2001; Vinu V Das, 2009; Xin Longbiao and Liu Chunlei. 2008 and Moonsuk Choi. 2008. Some existing meter reading techniques are: Touch Technology AMR, Radio Frequency AMR (Handheld, Mobile or "Drive-by" meter reading and Fixed Network AMR), Power Line Communication AMR, Wi –Fi.

With touch based AMR (http://www.electricity-today.com/et/issue0707/AMR_Evolving_Techology) a meter reader carries a handheld computer or data collection device with a wand or probe. The device automatically collects the readings from a meter by touching or placing the read probe in close proximity to a reading coil enclosed in the touchpad. When a button is pressed, the probe sends an interrogate signal to the touch module to collect the meter reading. The software in the device matches the serial number to one in the route database, and saves the meter reading for later download to a billing or data collection computer. As the meter reader has to go to the site of the meter, so this is not changing the current situation. Radio frequency based AMR (Rob Kavet and Gabor Mezei. 2010) can take many forms. The more common ones are Handheld, Mobile, and Fixed network. There are both two-way RF systems and one-way RF systems in use that use both licensed and unlicensed RF bands. Handheld AMR is where a meter reader carries a handheld computer with a built-in or attached receiver/transceiver (radio frequency or touch) to collect meter readings from an AMR capable meter. Mobile or "Drive-by" meter reading is where a reading device is installed in a vehicle. The meter reader drives the vehicle while the reading device automatically collects the meter readings. Fixed Network AMR is a method where a network is permanently installed to capture meter readings. This method can consist of a series of antennas, towers, collectors, repeaters or other permanently installed infrastructure to collect transmissions of meter readings from AMR capable meters and get the data to a central computer without a person in the field to collect it. Power line communication is very popular in AMR field (Agustín Zaballos *et al.,* 2009; John Newbury and William Miller, 2001; Xin Longbiao and Liu Chunlei, 2008; Xin Longbiao and Liu Chunlei, 2008 and Moonsuk Choi, 2008). Power line (PLC) AMR is a method where electronic data is transmitted over power lines back to the substation, then relayed to a central computer in the utility's main office. This would be considered a type of fixed network system the network being the distribution network which the utility has built and maintains to deliver electric power. Such systems are primarily used for electric meter reading. Some providers have interfaced gas and water meters to feed into a PLC type system.

All of the above technology have some limitation and is not feasible for Bangladesh perspective. In the touch technology, handheld technology and in mobile technology still the meter readers have to go to the houses, offices and other places where the meters are placed. So still the meter reader person is required. In addition we need extra devices which are very expensive. As a result they are not cost feasible. The PLC technology is not also feasible for Bangladesh perspective. In Bangladesh high voltages transmits through the power line cable. As the voltage is high so the transmitted data will be corrupted by the attenuation. All the power line cable of our country is not placed under the ground. It situated in the open air. So the cable faces different environmental problems. So the actual data may not transmit to the provider end. As a result this technology is also not feasible in our country. The fixed RF technology has small coverage area. As a result, this method consist of a number of series of antennas, towers, collectors, repeaters, or other permanently installed infrastructure to collect transmissions of meter readings from AMR capable meters. So this is not cost efficient for the customers. The main objective of this paper is to introduce an AMR system which is cost efficient. Some other important objectives are: i. to reduce data collection costs, ii. To improve meter reading accuracy, iii. to enable faster, more efficient reading times and billing process, iv. Significantly increase operational efficiency by providing real time pricing and time-of-use metering, v. to improve customer service and vi. enable conservation of resources.

## METHODOLOGY

In this research we analyzed the existing meter reading system of our country and found out the different problems of the present system. We studied the different technologies available in the world to reduce the meter reading problems and modeled out a feasible solution for our country. We have visited different areas of Dhaka city and interviewed the inhabitants of the areas. A number of web site and research papers have been studied to produce a good and feasible model for Bangladesh perspective. It was 8 months long research. The study was conducted at the Electrical Circuit Laboratory, American International University, Dhaka, Bangladesh during October 2009 to November 2010.

## EXPERIMENTS AND FINDINGS

### PROPOSED MODEL

Almost all the meter reading systems consists of three primary components. We divided the whole AMR system into four basic units. These are: Reading unit, Communication unit, Data receiving and processing unit, billing system.

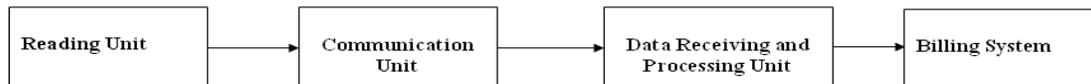

Fig. 1. Block Diagram of whole AMR system.

**Reading unit**

In this part basically two important jobs have been done. At first the analog meter reading was converted to digital bits sequence (0 or 1). After that the data are available in the microcontroller for transmission. First challenge is how we can get reading automatically from analog digital watt meter. It was analyzed that the rotations of the disk of the meter are needed to be counted. If it is possible to measure the number of rotation of the disk of the meter then the meter reading can be calculated. There are various types of sensors are available. Infrared can be used as a sensor. The infrared transmitter generates frequency and the receiver receives it. If there is no obstacle between the infrared transmitter and the infrared receiver than the infrared receiver give one value. But if there is any obstacle then the receiver gives another value. So this event can be used to count the rotation of the disk. The infrared transmitter and receiver have to place in such a way so that when the disk of the meter rotates it can recognize. A small hole is needed in the disk of the meter and the infrared transmitter and receiver should be placed in two opposite side of the disk of meter. They should be placed very close to each other so that when the hole cross the infrared then the transmitter and receiver can communicate. As a result when the disk is rotating the infrared can't communicate, but when the hole will come in between the infrared transmitter and the infrared receiver then they can communicate with each other. So a signal can be found for each time just after passing the hole which indicating that the disk is rotated by one. To control the infrared microcontroller is used i.e. required voltage for the infrared was supplied from the microcontroller. A microcontroller instruction has been written for the infrared transmitter to transmit the frequency and made the infrared receiver active by giving high signal. A circuit diagram was used to implement this. An application program has been written and burnt it in the microcontroller. A LED display was used which indicated that the infrared transmitter and receiver is working or not. If the communication established i.e. the hole found between the infrared transmitter and receiver then the LED will turned on and it will turn off if the hole is not found means there is no communication between the transmitter and receiver. To implement the reading unit some electrical components like infrared sensors (transmitter, receiver), resistors, Light Emitting Diode (LED) were used. For proper operation of the IR transmitter a 220 ohm resistor was connected. And the VDD was supplied to the IR transmitter through the pin7 (P7) of the microcontroller. Another part of the IR Transmitter was connected with VSS. The IR Receiver contains three PIN and was connected with VDD, VSS and the pin8 of the microcontroller as like the figure.

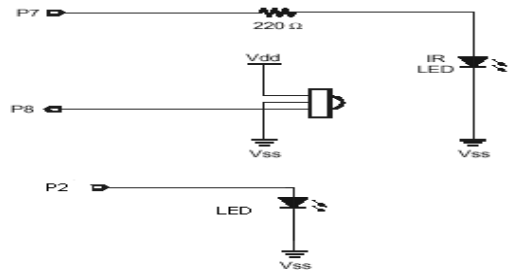

Fig. 2. Circuit diagram of reading units.

When the hole came between the infrared transmitter and receiver then the value of pin8 become 0 (zero). If the value is zero then a counter was incremented by one. Depending on the meter the unit can be calculated from the value of the counter. As a result the unit is stored in the microcontroller and is available for transmission. An example of the implementation picture is given in fig. 3.

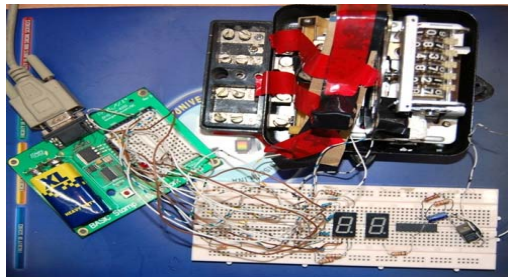

Fig. 3. Sample Implementation.

So the basic steps for taking the meter reading are as follows:

At first made a small hole in the disk of current existing meter -> Place the infrared sensor in specified way -> Check the hole -> If hole found increment the counter -> And write the data in the EPROM of microcontroller. The flow chart will be like this:

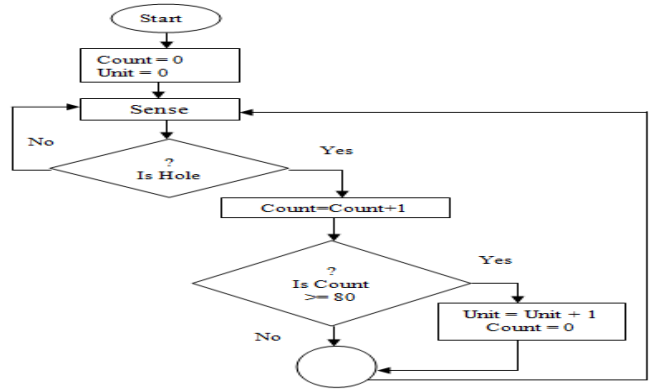

Fig. 4. Sample Flow chart of the basic steps of reading unit.

**Communication Unit**

This is one of the most important and challenging part of this system. This part is challenging in that sense, data is the most valuable part for meter reading and billing system. Data should be transmitted in an efficient manner without any loss of data. Let's describe about this challenging part. From the above description it is clear that digital data is always ready for transmission. Meter reading are stored in Microcontroller's EPROM and this data is always ready for transmission. But the main concern was how data can be transmitted efficiently? From the background study it is realized that all the existing communication units are not feasible for higher cost and infrastructure of Bangladesh. After studying different technologies, Wimax has been chosen for communication. In Wimax possibilities of data corrupting is very less. The coverage area is very high in Wimax and it is 10 km for NLOS communication. For the purpose of communication between meter end and the server end, a small miniature and low cost Wimax Transceiver module is required in each meter and in the server end. A transceiver module is a module which can transmit and receive data at a time. After searching a miniature and low cost transceiver module has been found.

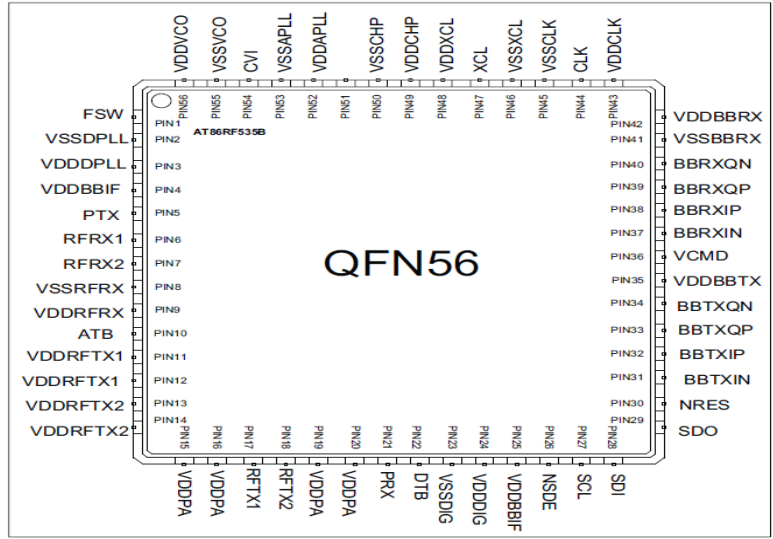

Fig. 5. Pin Information of AT86RF525B.

The model name is AT86RF535B. AT86RF535B is a fully integrated, low cost RF 3.5GHz Low-IF/Zero-IF conversion transceiver for WiMAX applications. It combines excellent RF performance, small size, and low current consumption. The AT86RF535 chip is fabricated on the advanced SiGe BiCMOS process AT46000. The transceiver combines LNA, PA driver, RX/TX mixer, RX/TX filters, VCO, Synthesizer, RX Gain control, and TX Power control, all fully digitally controlled. This transceiver module is miniature in size. This Wimax transceiver module can be set up inside of the current analog wattmeter. And this transceiver also cost very lower than other transceiver. In the server end there will be a transceiver and each meter will contain a miniature transceiver. A computer application will run at the server end which can send an address of a particular meter to the microcontroller and the microcontroller will supply the address to the transceiver. Then the transceiver will send the address to all meters. Generally all the transceiver of the meters will be in sleep mode. When a transceiver of the meter receives the address sent by the server transceiver then it compares that is the request is for itself or not. If the request is for itself then it give a high signal to the microcontroller and the microcontroller send the data to the transceiver. After getting the data from the microcontroller the transceiver transmit the data. The server end transceiver receives the data and provides the data to the server end microcontroller. The microcontroller sends the data to the computer.

This is the overall communication part of our system. Fig. 6 shows the data flow in communication unit of our proposed system.

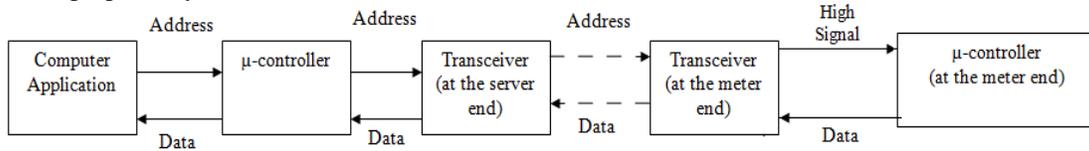

Fig. 6. Data flow in the communication units.

**Data receiving and processing unit**

This is the third part of the proposed AMR system. In this part the received data is processed by the system for future purpose. For data processing purpose a computer application has been developed. The task of the application was to take a meter number form the user and give the address to the microcontroller through serial port. Then the microcontroller does the communication task. After communication part the microcontroller get the data form transceiver and the meter reading is available in the server end microcontroller. Then the data is sent to the computer and the computer application receives the data from the microcontroller. This data can be stored in the database and can be displayed to the requested user.

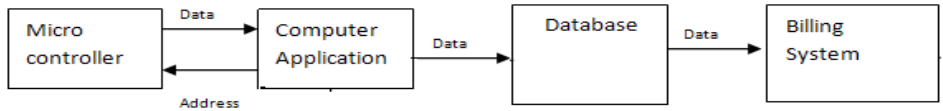

Fig. 7. Block diagram of data receiving and processing unit.

**Billing System**

The billing system has been developed in our system which can take the meter number and can generate bill for that meter. It uses the data of the database those are collected from the meter reading through all the unit of our system. This system also can be used for analysis on electricity usage for each meter.

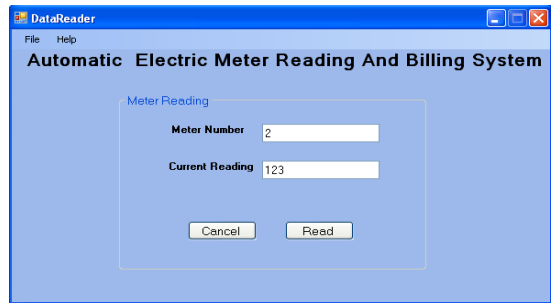

Fig. 8. Example on demand Reading using the computer application.

**Overall Conceptual design**

In this system the existing analog meter will be used and our proposed miniature module will be added to each meter. A module will be situated in the server end and this module will be connected with the server computer. The entire module will be connected through Wi-Max technology. The overall conceptual design has given in fig. 9.

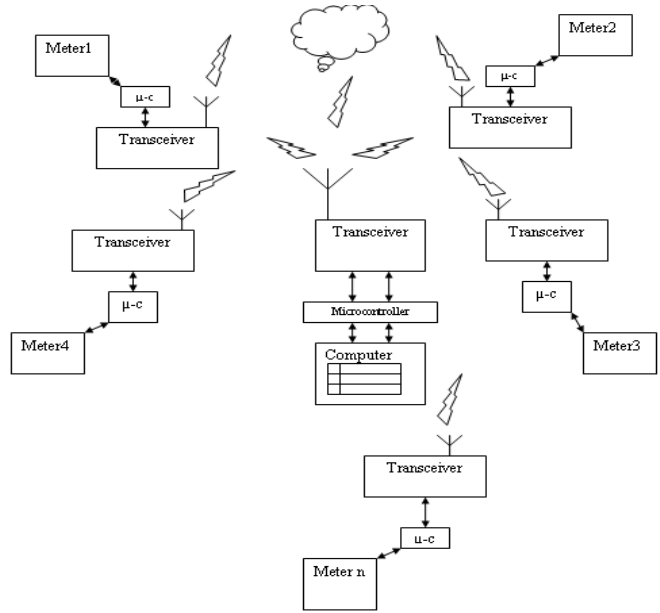

Fig. 9. Conceptual Diagram of our proposed AMR.

**WIMAX AS A TRANSMISSION MEDIA**

WiMAX, the Worldwide Interoperability for Microwave Access, is a telecommunications technology aimed at providing wireless data over long distances in a variety of ways, from point-to-point links to full mobile cellular type access. It is based on the IEEE 802.16 standard, which is also called Wireless MAN. The name WiMAX was created by the WiMAX Forum, which was formed in June 2001 to promote conformance and interoperability of the standard ( http://en.wikipedia.org/wiki/WiMAX ).

WiMAX has a theoretical maximum bandwidth of 75Mbps. This bandwidth can be achieved using 64QAM 3/4 modulation. 64QAM can only be utilized under optimal transmission conditions. WiMAX supports the use of a wide range of modulation algorithms to enable the most bandwidth to be realized under all conditions (http://en.wikipedia.org/wiki/WiMAX). WiMAX has a theoretical maximum range of 31 miles with a direct line of sight. Near-line-of-sight (NLOS) conditions will seriously limit the potential range. In addition, some of the frequencies utilized by WiMAX are subject to interference from rainfade. The unlicensed WiMAX frequencies are subject to RF interference from competing technologies and competing WiMAX networks. WiMAX can be used for wireless networking in much the same way as the more common WiFi protocol. WiMAX is a second-generation protocol that allows for more efficient bandwidth use, interference avoidance, and is intended to allow higher data rates over longer distances.

Here is a summary of comparison:

Table 1. Comparison of different wireless technology.

| Communication Technologies | PLC | UHF RADIO | Wi-Max |
|---|---|---|---|
| Data rate per connection | 3.0 Mbps | 9.5 Mbps | 75 Mbps |
| Range | - | 10 km | 50 km |

From the above table it can be understood that AMR with Wimax has better data rate and largest coverage area. If AMR using RF is used a large number of networking tools and resources will be needed. So the possibility of corrupting data is increased which is not feasible. We made a general consideration about Dhaka city. Over 20 lakhs household meter are available in Dhaka city. As the area of Dhaka city is not a big one so only one server is enough to coverage the whole Dhaka city. In addition here the corruption of data will be less. So in cost and data both senses Wimax is the appropriate wireless technology for AMR in Bangladesh.

**BENEFITS AND FUTURE WORK**

Tangible benefits: In this system no meter reader is required. By using this system the real time meter reading is possible. Some other tangible benefits of our system are: Reduce the cost of meter reading with our low cost module, Detect meter tempering, No need to change current existing meter, Miss use of electricity will be decreased, Reduce of corruption done by the meter reader, Need based system, Monitoring the meter reading, Feasible for Bangladesh etc. As the system is automated, the retrieved data can be used for other purpose like online billing system, demand analysis, other analysis etc. Finally we can say that the service provider will get the actual bill.

Intangible benefits: This system has some very important intangible benefits. As the service provider will get the actual bill, they will be satisfied by this system. The miss use of electricity will be decreased which will result reduced load shading. As the corruption is decreasing, the government will be benefited by this system. There will be no fake bill, so the customer will be satisfied. For future advancement more research can be done to develop meter reading for water meter and other types of meter reading.

**CONCLUSION**

Automatic meter reading using Wi-Max technology is an excellent idea for Bangladesh perspective. We tried proposed a cost efficient AMR system. We also researched to make a feasible system for Bangladesh perspective. If it is implemented in our country the corruption of electricity sector will be reduced as well as the loss will also be reduced. So Bangladesh government should take necessary steps to implement Automatic meter reading system in our country.